\documentclass[
    ,final            
  ]
  {aipproc}

\layoutstyle{8x11double}

\usepackage{color}

\begin{document}

\title{Classification of Topological Insulators and Superconductors}




\author{Andreas P. Schnyder}{
  address={Kavli Institute for Theoretical Physics, University of California, Santa Barbara, CA 93106, USA} 
}

\author{Shinsei Ryu}{
  address={Department of Physics, University of California, Berkeley, CA 94720, USA} 
}

\author{Akira Furusaki}{
  address={Condensed Matter Theory Laboratory, RIKEN, Wako, Saitama 351-0198, Japan} 
}

\author{Andreas W. W. Ludwig} {
  address={Department of Physics, University of California, Santa Barbara, CA 93106, USA}
}

%
%

\begin{abstract}
An exhaustive classification scheme of topological insulators and 
superconductors is presented.
The key property of topological insulators (superconductors) is the appearance 
of gapless degrees of freedom
at the interface/boundary between a topologically trivial and a topologically 
non-trivial state.
Our approach consists in reducing the problem of classifying 
topological insulators (superconductors) in $d$ spatial dimensions to the 
problem of Anderson localization at a $(d-1)$ dimensional boundary of
the system.
We find that in 
each spatial dimension there are precisely 
five
distinct
classes
of topological insulators (superconductors).
The different topological sectors within a given topological 
insulator (superconductor) can be labeled by an integer winding number  
or a $Z_2$ quantity. One of the five topological insulators is
the ``quantum spin Hall'' (or: $Z_2$ topological) insulator 
in $d=2$, and its generalization in $d=3$ dimensions.
For each dimension $d$, the five topological insulators correspond
to a certain subset of five of the ten generic symmetry classes of
Hamiltonians introduced more than a decade ago by Altland and Zirnbauer
in the context of disordered systems (which generalizes the three well known 
``Wigner-Dyson'' symmetry classes).
\end{abstract}

\maketitle




\section{A.  Introduction}

\begin{figure}
\includegraphics[height=.275\textwidth]{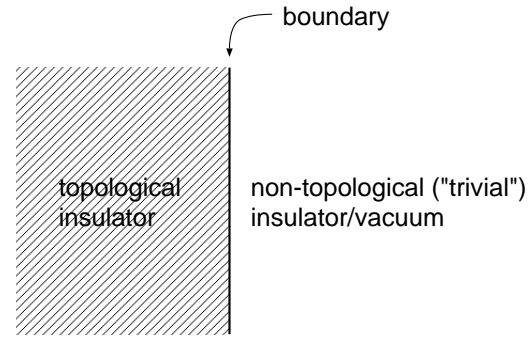}
\caption{Interface between a topological, and a topologically trivial insulator.}
\label{Figure1}
\end{figure}

We will
give a review of the (exhaustive) classification scheme of
{\it topological insulators 
(or: superconductors)}\footnote{
In this work we  consider topological insulators (superconductors) 
without interactions.
Since these  are gapped states in the $d$-dimensional bulk, 
such states will be stable to sufficiently weak interactions.
However, under what conditions certain different topological states
are adiabatically connected when interactions are included is,
to a large extent, an open problem to-date.
}
presented  in Ref.~\cite{SchnyderRyuFurusakiLudwig2008}.
We can think of
topological insulators (superconductors)
as being gapped states (thus ``insulators'')
in $d$ spatial dimensions (we consider here  $d=1,2,3$) with the following
property:  if we terminate the topological 
insulator (superconductor)
against a ``topologically trivial'' state, such as e.g. simply vacuum,
gapless degrees of freedom {\it will necessarily appear} 
at the interface (``boundary'') between the topologically 
trivial and the topologically non-trivial states -- see Fig.~\ref{Figure1}. 
(We will present some simple,
and well known examples shortly.) Moreover, the so-appearing
gapless boundary degrees of freedom
are completely robust to perturbations.
For example, we may subject
the topological insulator (superconductor)
to arbitrary random potentials or perturbations 
no matter how strong, 
without destroying the ``gaplessness'' of the boundary degrees of freedom,
as long as these perturbations 
do not close the bulk gap and
preserve the generic symmetries of the system 
(what is meant by these symmetries will be made precise below, and is indeed
of fundamental importance in our work). Clearly, these gapless modes must be of a
very special kind, since typically, gapless degrees of freedom
tend to  become localized in the presence of random potentials, certainly
if the latter are sufficiently strong (this is the phenomenon of
``Anderson localization'' for non-interacting systems).

In short, the approach used in this work to classify
topological insulators (superconductors)
in $d$ spatial dimensions 
consists in classifying gapless systems of fermions
(corresponding to the boundary degrees of freedom) 
which cannot be localized by disorder.
Thus, we reduce the problem of classifying topological insulators (superconductors)
in $d$ spatial dimensions to a problem of Anderson localization in $(d-1)$ 
dimensions.
In this work we solve this problem of
Anderson localization, and thereby the classification problem for topological 
insulators (superconductors).

Topological insulators (superconductors) are inherently ``holographic'' states:
the nature of the $d$-dimensional gapped topological bulk state can be
read off from the (holographic) ``image'' or ``shadow''
of these topological properties
on the system's boundaries. Indeed, there is a one-to-one correspondence
between the topological properties of the gapped bulk and properties of the
gapless surface degrees of freedom. These notions
are of course familiar from the quantum Hall 
effect,\footnote{
See e.g.\ \cite{WittenChernSimons,MooreRead,Hatsugai}; see also
\cite{WittenChernSimonsGravity} for a different context.}
and it will be
useful to remind the reader of (simple) well known examples of such
quantum states.

\begin{figure}
  \includegraphics[height=.22\textwidth]{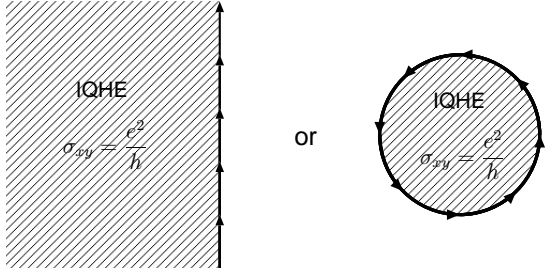}  
  \caption{Chiral edge states of the integer quantum Hall insulators.
}
\label{Figure2iqhe}
\end{figure}

{\it Well known examples of topological insulators (superconductors):}

(i): The probably best known example of a topological insulator is
the {\it integer quantum Hall insulator} 
(i.e., a filled Landau level).
In this non-interacting $d=2$  electron system 
time-reversal symmetry  (TRS)
is broken due to the applied magnetic field. If we terminate the quantum Hall
insulator by a one-dimensional boundary 
against ``vacuum'', a gapless edge state is known to appear
(see Fig.~\ref{Figure2iqhe}).
This edge state possesses a chirality inherited from the applied
magnetic field (broken TRS),
and propagates only in one direction;  therefore it cannot be localized by
disorder.

(ii): Another example in $d=2$ is the chiral $p_x+i p_y$  superconductor
(see e.g.\ \cite{ReadGreen}).
This is a gapped
superconductor, which also breaks TRS. The non-interacting system in question
is the system of quasi-particle excitations deep inside the superconducting state,
as described by the Bogoliubov-de Gennes (BdG) equation. This is an example of
a topological superconductor, as can be seen by terminating the chiral $p_x + i p_y$
state against vacuum (or an otherwise structureless 
``standard'' superconducting state): 
again,
at the interface (boundary) a chiral
edge mode is known to occur, which, since it propagates only in
one direction cannot be localized by disorder [just as that of example (i)].
However, since charge is not a conserved quantity in a superconductor,
this chiral edge mode only transports heat (energy) and not charge.
This makes clear that while also being a topological gapped state in $d=2$,
the chiral $p_x+i p_y$  superconductor 
possesses different ``symmetries'' than the $d=2$ integer quantum Hall state
in example (i).
(This notion of ``symmetries'' will be made precise below.)

\begin{figure}
  \includegraphics[height=.2\textwidth]{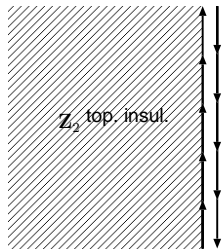}
  \caption{Chiral edge states of the $Z_2$ topological insulator
 (``quantum spin Hall effect'') 
in $d=2$ dimensions.}
\label{Figure3qshe}
\end{figure}

(iii): Another topological insulating state, often referred to
as the $Z_2$-topological
insulator~\cite{ZtwoTopologicalInsulators,ReviewKonigZhang}, or 
the ``quantum spin Hall'' (QSH) state, has recently attracted much attention. 
This state is known to exist in $d=2$ and in $d=3$ dimensions and,
as opposed to the previous two examples, does not break TRS. 
It is known to occur in certain band insulators
with strong spin-orbit interactions.
Let us first discuss the $d=2$ case, realized experimentally 
e.g.\ in HgTe/(Hg,Ce)Te semiconductor
quantum wells~\cite{RefHgTe}.
Because TRS is not broken, it is not as obvious as in examples (i) and (ii)
why the gapless boundary degrees of freedom appearing at the interface
terminating  the $d=2$ $Z_2$-topological insulator against vacuum
cannot be localized by disorder. However, this edge state consists
of a {\it single} Kramers doublet corresponding to a  {\it single}
pair of modes propagating in opposite
directions (see Fig.~\ref{Figure3qshe}),
which cannot be mixed by any TRS impurity potential.\footnote{Indeed,
a one-dimensional extended (not localizing) state
had already  been observed~\cite{Zirnbauer1992}
in studies of (quasi 1D) Anderson localization problems
with spin-orbit scattering in 1992,  but this
observation was not understood until recently:
truly quasi one-dimensional systems
with spin-orbit scattering 
must always possess {\it two} of Kramers doublets which indeed are not protected
from localization by disorder; however, when the one-dimensional system
is the boundary of what is known today as a two-dimensional $Z_2$-topological
insulator, a pair of edge states which form a {\it single} Kramers doublet
appears on each boundary, 
and such a pair evades Anderson localization.}
The $Z_2$-topological insulator~\cite{ZtwoTopologicalInsulators}, or 
the QSH state, is also known to exist in $d=3$
dimensions~\cite{REF3DZ2TopIns-Moore,REF3DZ2TopIns-Roy,REF3DZ2TopIns-Fu}.
It is realized in Bismuth-Antimony alloys, as demonstrated in recent
experiments~\cite{HassanPrincetonExptsOne,HassanPrincetonExptsTwo}.

In this work, we ask ourselves the question: which $d$-dimensional
(non-interacting) fermion
systems possess gapped ground
states with topologically non-trivial properties, 
i.e., which systems are  ``topological insulators (superconductors)'',
as described above?
How many ``different'' such systems are there? Are there infinitely many, or only
a finite number of them? How do these properties depend on the spatial dimensionality $d$?
Is there any systematics underlying these different systems?

The answer to these questions  turns out to be both  deep and interesting:
in every spatial dimension $d=1,2,3$ there exist {\it precisely}
five different classes of topological insulators (superconductors). 
In $d=3$ dimensions, for example, there exist besides the
$Z_2$-topological insulator discussed in example (iii) above,
four more, and all five topologically non-trivial states possess
TRS (in $d=3$). Our results
are summarized in Table \ref{table2}.
In order to explain Table \ref{table2}, we  first need to explain the very
general symmetry classification of quantum mechanical Hamiltonians,
and explain why this is a fundamental concept underlying the classification scheme of
topological insulators (superconductors).
This will be done in the following section, Section B.
In section C we will explain with a few very simple examples
how topology can arise in simple systems such as band insulators.
In section D we will describe the classification of topological
insulators (superconductors) in $d=3$ spatial dimensions.
Sections E and F will provide a summary of the same classification
in $d=2$ and $d=1$ dimensions. Section G provides  a discussion
and concluding remarks.

\begin{table}
\label{table1}
\begin{tabular}{|c|c||c|c|c||c|c|}
\hline
System  &  Cartan nomenclature & TRS&PHS &SLS & Hamiltonian    & NLSM
 (ferm. replicas)\\\hline\hline
standard      &A (unitary)    &$0$ &$0$&$0$&$\scriptstyle U(N)$&$\scriptstyle U(2n)/U(n)\times U(n)$ \\ \cline{2-7}
(Wigner-Dyson)&AI (orthogonal)&$+1$&$0$&$0$&$\scriptstyle U(N)/O(N)$&
 $\scriptstyle Sp(2n)/Sp(n)\times Sp(n)$ \\ \cline{2-7}
           &AII (symplectic)&$-1$&$0$&$0$&$\scriptstyle U(2N)/Sp(2N)$&$\scriptstyle O(2n)/O(n)\times O(n)$\\\hline\hline
chiral        &AIII (chiral unit.) &$0$&$0$&$1$&${\scriptstyle U(N+M)/U(N)\times U(M)}$ &$\scriptstyle U(n)$\\ \cline{ 2-7}
(sublattice)&BDI (chiral orthog.)&$+1$&$+1$&$1$&$\scriptstyle SO(N+M)/SO(N)\times SO(M)$& $\scriptstyle U(2n)/Sp(n)$ \\ \cline{2-7}
   &CII (chiral sympl.) &$-1$&$-1$&$1$&$\scriptstyle Sp(2N+2M)/Sp(2N)\times Sp(2M)$& $\scriptstyle U(2n)/O(2n)$ \\\hline \hline
BdG           &D                        &$0$ &$+1$&$0$ &$\scriptstyle SO(2N)$      &$\scriptstyle O(2n)/U(n)$\\ \cline{2-7}
              &C                        &$0$ &$-1$&$0$ &$\scriptstyle Sp(2N)$    & $\scriptstyle Sp(n)/U(n)$ \\ \cline{2-7}
              &DIII                     &$-1$&$+1$&$1$ &$\scriptstyle SO(2N)/U(N)$    & $\scriptstyle O(2n)$ \\ \cline{2-7}
              &CI                       &$+1$&$-1$&$1$ &$\scriptstyle Sp(2N)/U(N)$       & $\scriptstyle Sp(n)$\\ \hline
\end{tabular}
\caption{
Ten symmetry classes of single particle Hamiltonians classified in terms of
the presence or absence of time-reversal symmetry (TRS) and
particle-hole symmetry (PHS), as well as  
sublattice (or ``chiral'') symmetry
(SLS)~\cite{ZirnbauerJMathPhys1996,AltlandZirnbauerPRB1997,HuckleberryEtAl}.
In the table, the absence of symmetries is denoted  by ``$0$''.
The presence of these symmetries is denoted either by ``$+1$'' or ``$-1$'',
depending on whether the (anti-unitary) operator 
implementing the symmetry squares to ``$+1$''or ``$-1$''.  
For the first six entries of the table
(which can be realized in non-superconducting systems) 
TRS $=+1$ when the SU(2) spin is integer 
and TRS $=-1$ when it is a half-integer.
For the last four entries, the superconductor ``Bogoliubov-de Gennes'' (BdG)
symmetry classes 
(denoted by the symbols D, C, DIII, and CI in ``Cartan nomenclature''),
it turns out that the Hamiltonian preserves SU(2) spin-1/2 rotation
symmetry when PHS=$-1$ 
whilst it does not preserve SU(2) when PHS=$+1$. 
The column entitled ``Hamiltonian'' lists the spaces to which the quantum
mechanical
time-evolution operators of each symmetry class belong (see section B).
The column entitled ``NLSM (ferm. replicas)'' lists the ``target spaces''
of Non-Linear Sigma Model
field theories describing Anderson localization physics in each symmetry class 
(see section B).
}
\end{table}

Our work~\cite{SchnyderRyuFurusakiLudwig2008}
demonstrates an unexpected relationship between two apparently
rather unrelated subjects.   One is the symmetry classification of general
quantum mechanical Hamiltonians (``The 10-fold Way'' to be reviewed in Section B below),
which is at the root of theories of disordered systems, be it Anderson localization,
or random matrix theory.\footnote{
This classification appeared in the work of 
Zirnbauer~\cite{ZirnbauerJMathPhys1996}, and Altland and 
Zirnbauer~\cite{AltlandZirnbauerPRB1997,HuckleberryEtAl},
more than a decade ago. 
It extends the well familiar ``Wigner Dyson'' classification of Hamiltonians
(``unitary, orthogonal, symplectic'' classes).}
The other is the classification of topological insulators.
It is rather surprising, initially, that
a relationship exists between these two subjects. Moreover, a relationship
with a third subject has recently emerged.  Recent work by A.~Kitaev
reached~\cite{KitaevLandau100Proceedings}, by using K-theory,
the same conclusions as those obtained in~\cite{SchnyderRyuFurusakiLudwig2008}
for the classification of topological insulators.
Thus there are remarkable connections between the seemingly rather disjoint
subjects  of (i)  topological insulators, (ii) Anderson localization, random systems,
and random matrix theory,
as well as (iii)  K-theory.

\section{B.  Symmetry Classification of Hamiltonians -- ``The 10-fold Way''}

Consider the gapped Hamiltonian and the corresponding ground state
of a $d$-dimensional topological insulator.\footnote{Consider the
BdG Hamiltonian in the case of the topological superconductor.}
Because of the presence of the gap, we may deform the Hamiltonian slightly,
by adding various perturbations to it
(which preserve the generic symmetries such as, e.g., time reversal symmetry),
while still preserving the gap.
In this way, we map out an entire gapped phase. We may then ask the question:
how many different such phases can a system possess, so that in going from one
phase to another, a quantum phase transition has to be crossed? Now,
clearly, because of the bulk gap, we will also remain in the same phase by
perturbing our Hamiltonian by  perturbations which break translational
invariance (certainly, as long as these perturbations are small enough).
The most general such perturbation is what we call a  random perturbation. 
The perturbed
Hamiltonian is thus a random (=lacking translational invariance) 
gapped Hamiltonian. If the original (unperturbed)
phase was topological (in the sense described above), then the random
Hamiltonian will be in the same (topological) phase. Therefore we see
that in attempting to classify topological phases, we need to consider
in general random gapped  Hamiltonians. Thus a given gapped topological phase
is associated with a certain class of gapped random Hamiltonians.
Hence we are led to study the classification of
random gapped Hamiltonians.
How many such Hamiltonians are there? Clearly, in attempting to classify
random Hamiltonians, we can only use the ``{\it most generic quantum mechanical
symmetries}'', translational invariance not being one  of them.
The symmetry properties that every quantum mechanical Hamiltonian
can be classified by are time-reversal symmetry  (TRS) and charge-conjugation
(or: ``particle-hole'') symmetry (PHS). Investigating the properties of
a general Hamiltonian under such symmetries yields the now famous ten symmetry
classes (the ``ten-fold way''), originally described in the seminal
work of Zirnbauer~\cite{ZirnbauerJMathPhys1996},
and Altland and Zirnbauer~\cite{AltlandZirnbauerPRB1997,HuckleberryEtAl},
more than a decade ago. 
This classification extends and completes the familiar ``three-fold way''
classification scheme of Wigner and Dyson~\cite{RefWignerDyson},
going back to the origins of random matrix theory and the study of complex nuclei.
The reason why there are only ten possible symmetry classes of quantum
mechanical Hamiltonians is easy to understand by considering TRS and PHS.
Let us begin with the  time-reversal  operator $T$ which is an anti-unitary operator.
Thus $T$ is the product of a unitary operator $U_T$ and the ``complex conjugation operator''
$K$, i.e., $T= K U_T$.
On the first quantized Hamiltonian ${\cal H}$
time reversal acts as ${\cal H} \to$ $U_T^\dagger {\cal H}^* U_T$. Now,
any Hamiltonian can behave in three possible ways under TRS: 
(i): it is not invariant under TRS, which case we denote by TRS$=0$,
or, (ii):  
it is invariant under TRS and the (anti-unitary) time reversal operator
$T$ squares to $+1$, which case we denote by TRS$=+1$,
or, (iii)
it is invariant under TRS and the (anti-unitary) time reversal operator
$T$ squares to $-1$, which case we denote by TRS$=-1$.
Similarly, it is possible
to describe the particle-hole (charge-conjugation) symmetry (PHS) 
operator
as an anti-unitary operator $C$, when acting on a non-interacting system
(see Ref.~\cite{SchnyderRyuFurusakiLudwig2008} for details).
Therefore, analogously, the possible behaviors of any (non-interacting)
Hamiltonian under PHS
is PHS$= 0, +1, -1$ (meaning that PHS  
is not a symmetry, or  is a symmetry  and the anti-unitary operator $C$
squares to $+1$ or $-1$, respectively).
It is now easy to see that there are precisely
ten symmetry classes (i.e., those found by Zirnbauer and 
Altland \cite{ZirnbauerJMathPhys1996,AltlandZirnbauerPRB1997,HuckleberryEtAl}):
There are $3 \times 3=9$ different choices for the behavior of any Hamiltonian
under TRS and PHS. A moment's thought shows that for $8$ of these $9$ choices
the behavior of the Hamiltonian under the product%
\footnote{The so-defined
symmetry operation  is sometimes called ``${\rm \bf s}$ub${\rm \bf l}$attice 
${\rm \bf s}$ymmetry'' (or also: ``chiral symmetry''), hence
the notation SLS, because a particular (and popular) example of
this symmetry arises in
systems described by a hopping Hamiltonian on a bipartite lattice, where only
matrix elements for hopping between the two different sublattices of the
bipartite lattice are non-vanishing. However, this symmetry can be viewed 
generally simply as the product of $T$ and $C$, as stated;
we still denote it by the symbol SLS.} 
SLS $:= T*C$ of TRS and PHS, which is a unitary operator, is uniquely fixed. 
(We write SLS$=0$ if the operation SLS is not a symmetry of the Hamiltonian,
and SLS$=1$ if it is.)
The only case when the behavior under the combined transformation SLS
is not uniquely 
determined by the behavior under TRS and PHS is
when TRS$=0$ and simultaneously PHS$=0$. In this case either SLS$=1$ or SLS$=0$
is possible. This reasoning gives hence $(3\times3-1)+2=10$ possible behaviors
of a Hamiltonian. 

These are the ten symmetry classes mentioned above,
which are listed in Table~\ref{table1}.
The column ``Hamiltonian'' describes the nature of the time evolution operator
$\exp\{ i t {\cal H} \}$, where ${\cal H}$ is the first quantized Hamiltonian.
If we consider a discretized version of the system, e.g., on a (finite) lattice
(as, e.g., for a hopping Hamiltonian), then the Hamiltonian ${\cal H}$
is a finite $N\times N$ matrix.\footnote{$N$ is the product of the number of
lattice sites, times the number of
spin orientations (e.g., spin-up and spin-down), if applicable, etc. ...}
In each symmetry class the time evolution operator is an element of
the particular group or symmetric space, listed in this column.
For example, if the system
has no symmetries at all, it belongs to symmetry class A. This is the case for
a quantum Hall system, where TRS is 
broken.\footnote{The quantum Hall insulator mentioned in example (i) above
belongs to symmetry class A.}
In this case, there are no (symmetry) constraints
on the Hamiltonian and ${\cal H}$ is a generic Hermitian matrix.
The time-evolution
operator is thus a generic unitary matrix (an element of the group 
$U(N)$ of unitary matrices,  as noted in Table \ref{table1}), 
without any further conditions imposed.
The first three rows in Table \ref{table1} denote thus
the standard (``Wigner-Dyson'') 
symmetry classes (``unitary, orthogonal, symplectic'');
these  are  distinguished only by the presence or absence of
TRS, and possess no other symmetries.
Example (iii) discussed above belongs to symmetry class AII, in which
the only symmetry is TRS, with the (non-unitary) TRS operator
squaring to minus the  identity operator.

The next three rows in Table \ref{table1} are identical to the first
three rows, except that all possess  an additional SLS symmetry (SLS$=1$),
whereas SLS$=0$ for the first three rows.
We will defer discussion of these symmetry classes for now, except that
we 
recall (footnote 6) that 
examples of simple realizations of all
three (AIII, BDI, CII) can be obtained from hopping models
where particles only hop between the two sublattices of a bipartite
lattice, with  the corresponding TRS properties also imposed.

The last four columns describe the symmetry properties of the fermionic
quasiparticles in  superconductors (or certain superfluids), 
deep inside the superconducting state, within a mean field
treatment of pairing.
Their dynamics is described by the BdG Hamiltonian, which is the Hamiltonian
${\cal H}$
whose properties are listed in the column entitled ``Hamiltonian''.
Any BdG Hamiltonian possesses by construction a PHS, as indicated in Table~\ref{table1}.
Example 
(ii) of the $d=2$ dimensional $p+ip$ superconductor 
(of spinless fermions)
belongs to symmetry class D (7th row):
the system possesses no symmetries (including TRS) 
other than the PHS inherent in all BdG Hamiltonians.
Let us conclude by pointing out that symmetry class 
DIII (9th row) describes the superfluid phase of $^3$He-B~\cite{volovikHe3B},
whereas the CI (10th row) describes, e.g.,  singlet superconductors.
Topological phases of the latter, in $d=3$ dimensions,
were discussed in the recent paper~\cite{SchnyderRyuLudwig2009arXiv}.

\begin{table}
\label{table2}
\begin{tabular}{|c|c||c|c|c||c|c|c|}
\hline
System         &  Cartan nomenclature & TRS & PHS & SLS & $d=1$     & $d=2$       & $d=3$           \\\hline\hline
standard       &  A (unitary)         & $0$ &$0$  & $0$ &  -        & ${\bf Z}$   &  -            \\ \cline{2-8}
(Wigner-Dyson) &AI (orthogonal)       &$+1$ &$0$  &$0$  &  -        & -           &  -               \\ \cline{2-8}
               &AII (symplectic)      &$-1$ &$0$  &$0$  &  -        &  ${\bf Z}_2$& ${\bf Z}_2$    \\\hline\hline
chiral         &AIII (chiral unit.)   &$0$  &$0$  &$1$  &${\bf Z}$  &  -          & ${\bf Z}$      \\ \cline{ 2-8}
(sublattice)   &BDI (chiral orthog.)  &$+1$ &$+1$ &$1$  &${\bf Z}$  & -           &        - \\ \cline{2-8}
               &CII (chiral sympl.)   &$-1$ &$-1$ &$1$  &${\bf Z}$  & -           & ${\bf Z}_2$   \\\hline \hline
BdG           &D                      &$0$  &$+1$ &$0$  &${\bf Z}_2$& ${\bf Z}$   & - \\ \cline{2-8}
              &C                      &$0$  &$-1$ &$0$  & -         &  ${\bf Z}$  & -        \\ \cline{2-8}
              &DIII                   &$-1$ &$+1$ &$1$  &${\bf Z}_2$&${\bf Z}_2$  & ${\bf Z}$ \\ \cline{2-8}
              &CI                     &$+1$ &$-1$ &$1$  & -   & -    & ${\bf Z}$ \\ \hline
\end{tabular}
\caption{
Summary of the {\it main result of this paper}:
listed are again the 
ten symmetry classes of single particle Hamiltonians (from TABLE 1) 
classified in terms of the presence or 
absence of time-reversal symmetry (TRS) and particle-hole symmetry (PHS), as well as  
sublattice (or ``chiral'') symmetry (SLS)~\cite{ZirnbauerJMathPhys1996,AltlandZirnbauerPRB1997,HuckleberryEtAl}.
The last three columns list all possible 
topologically non-trivial quantum ground states as a function of 
symmetry class 
and spatial dimension $d$.
The symbols ${\bf Z}$ and ${\bf Z}_2$ indicate
that the space of quantum ground states is partitioned into different topological sectors labeled by an 
integer (${\bf Z}$), or a ${\bf Z}_2$ quantity (two sectors only), respectively.
}
\end{table}

Let us finally comment briefly
on the last column of Table \ref{table1}. 
(An understanding of this  column is not 
required in order to be able to follow the rest of this review.)
It refers to the conventional long-wavelength 
description of Anderson localization 
of non-interacting fermions  subject to disorder potentials, in terms of 
a Non-Linear Sigma Model (NLSM)  field theory. 
A NLSM can be viewed as a generalization of the classical Heisenberg ferromagnet,
described by a model of classical unit vector spins. These spins can sweep out
a sphere, the simplest example of what is called a symmetric space. It is known
since the days of the mathematician E. Cartan, that there exist only 10 types
of symmetric spaces (barring so-called ``exceptional'' cases). In general, 
Anderson localization 
transitions
can be formulated in terms of  NLSM field theories of generalized spins, which
sweep out one of the 10 symmetric spaces (the ``target space''),
defining the NLSM.
The symmetry class determines which target space is to be used, and these
are listed\footnote{Specifically, we chose here the simplest formulation in
terms of a set of $n$ ``fermionic replicas'', where $n$ has to be taken to zero
at the end. In this formulation the symmetric spaces are all compact,
when $n$ is finite. 
The homotopy group of a symmetric space $\mathcal{M}$
tells us if it is possible to add a topological term ($\theta$ term when
$\Pi_d(\mathcal{M})={\bf Z}$ and $Z_2$ term when $\Pi_d(\mathcal{M})={\bf Z}_2$)
to a NLSM.
A technically better controlled, but equivalent
formulation can be provided using
a supersymmetric formulation~\cite{EfetovBook},
in which the manifolds listed in the last column
are replaced by supermanifolds~\cite{SupermanifoldsBryceDeWittBook}, 
containing compact, non-compact, and fermionic
coordinates.}
in the last column of Table~\ref{table1}.

\section{C. The origin of topology in band insulators}

\begin{figure}[t!]
  \includegraphics[height=.24\textwidth]{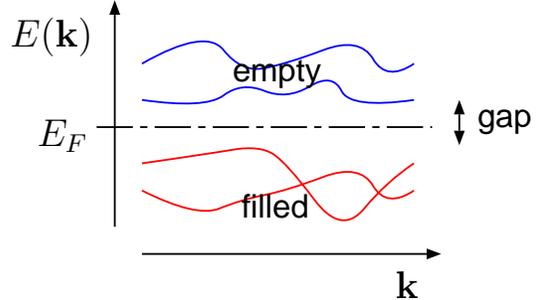}  
  \caption{Schematic band structure of a typical  band insulator.}
\label{Figure4bands}
\end{figure}

In order to illustrate in the simplest terms how topological properties
arise in topological insulators, let us begin with a translationally
invariant example. A topological insulator is a very simple system.
It is a band insulator of non-interacting fermions,
meaning that there is a gap between valence and conduction bands,
and the fermi level $E_F$ lies in this gap (see Fig.~\ref{Figure4bands}).
Due to the assumed translational invariance, the insulator is
described in momentum space by a matrix equation for every
value of momentum ${\vec k}$ in the Brillouin zone
$$
{\cal H}({\vec k}) \ \  \left| u_a(\vec k) \right\rangle
\ \ =
\ \
E_a({\vec k}) \ \ \left| u_a(\vec k) \right\rangle,
$$
where $a$ denotes an index labeling different bands.
Let us now consider, for every momentum ${\vec k}$ in
the Brillouin zone,  the projection operator onto the filled
(Bloch) states,
\begin{equation}
{P}({\vec k}) \ \  :=
\ \ 
\sum_{a}^{\rm filled}
\ \ 
\left| u_a(\vec k) \right\rangle 
\ \left\langle u_a(\vec k) \right|  .
\label{SpectralProjectorP}
\end{equation}
Instead of 
${P}({\vec k}) $ it turns out to be more convenient
to use the operator\footnote{ ${\bf 1}$ denotes the identity operator.}
\begin{equation}
{Q}({\vec k}) \ \  :=
{\bf 1}
- 
2 \ {P}({\vec k}) \ \ 
\label{HamiltonianQ},
\end{equation}
which has the following properties (as one readily checks)
$$
{Q}^{\dagger}({\vec k}) \ \  
= 
\ \  
{Q}({\vec k}),
\quad 
[{Q}^{\ }({\vec k})]^2 = {\bf 1},
\quad
\mathrm{tr}\,[
{Q}({\vec k})]
= m-n,
\label{ConditionsQk}
$$
where $m$ and $n$ denote the number of filled and empty bands,
respectively.
The Hermitian operator 
$
{Q}({\vec k}) 
$
plays the role of the 
Hamiltonian,
carrying only the essential information about the
insulator in question. It has eigenvalues $\pm 1$.
This ``simplified Hamiltonian'' is obtained from ${\cal H}({\vec k})$
by assigning, say,  to all occupied bands the energy $-1$ and to
all empty bands the energy $+1$
(while all wave functions remain unchanged).
Since we are only interested in the properties
of the phase described by the insulator, we may deform the actual 
Hamiltonian
of the band insulator until
it acquires the simple form
${Q}({\vec k})$, while remaining in the same phase.

In order to see how to use the ``Hamiltonian''
${Q}({\vec k})$, let us begin by considering
a  band insulator in the simplest symmetry class,
in which there are no conditions whatsoever imposed
on the Hamiltonian. This is symmetry class A, where the Hamiltonian
${\cal H}$ is nothing but a general Hermitian matrix.
In this symmetry class, the set of $n+m$ eigenvectors (each
being an $n+m$-dimensional vector) forms an arbitrary
unitary matrix, i.e., an element of $U(n+m)$.
There is however a simple ``gauge symmetry'', because
relabeling the empty and the filled states amongst
themselves (and forming arbitrary linear combinations amongst them)
does not change the physics. Therefore, the Hamiltonian 
${Q}({\vec k})$ is actually an element of the
so-called ``Grassmannian'' 
\label{Grassmannian}
\begin{equation}
{Q}({\vec k}) \in U(n+m)/[U(n)\times U(m)] .
\end{equation}
Since ${\vec k}$
runs over the Brillouin zone $BZ$, the
``Hamiltonian'' of the band insulator is
a map from the Brillouin zone into the Grassmannian,
$$
{Q}:
BZ \to  U(n+m)/[U(n)\times U(m)],
$$
\begin{equation}
{\vec k} \to 
{Q}({\vec k}) .
\qquad
\qquad
\qquad
\label{MapsBZGrassmannian}
\end{equation}
Let us summarize. The  
Hamiltonian 
of a band insulator can be continuously deformed
to the simple form
${Q}({\vec k})$
while remaining in the same phase (i.e., without crossing
a quantum phase transition).
Now, the question as to how many inequivalent phases
there are, amounts to asking how many
different maps
${Q}({\vec k})$ as in \eqref{MapsBZGrassmannian}
there are which cannot be continuously deformed into each other.
This question, on the other hand, is 
answered\footnote{
Besides the  features described by the homotopy
group (describing so-called ``strong topological'' insulators), 
there are additional  features related to the fact that the Brillouin
zone is a $d$-dimensional torus.
These are so-called ``weak topological'' features
(see \cite{REF3DZ2TopIns-Moore})
related to the presence of layers of topological insulators
\cite{KohmotoHalperin},
i.e., in one dimension less than the space dimension $d$.
So-called ``weak topological insulators'' possess only the latter,
by not the former topological features.
}
by the
homotopy group of the map in \eqref{MapsBZGrassmannian}.

Let us consider this in dimensions $d=2$ and $d=3$.

In $d=2$ the relevant homotopy group is 
$$
\Pi_{2} \left [
U(n+m)/[U(n)\times U(m)]
\right ]
= {\bf Z},
$$
where ${\bf Z}$ is the set of all integers.
This means that for every integer
there exists a band insulator in $d=2$ dimensions
in symmetry class A, and band insulators
corresponding to different integers cannot be continuously
deformed into each other without crossing a quantum phase
transition. We have encountered precisely these band insulators
already in example (i). These are the quantum Hall insulators,
and the integer characterizing the insulator denotes precisely
the number of chiral edge states. When the number of
edge states changes, a  quantum 
phase transition necessarily has to be crossed.
These are precisely the well studied quantum Hall
plateau transitions (driven by disorder).\footnote{The field theory describing
this transition is the $d=2$ dimensional NLSM on the target
space listed for class A in the last column of Table~\ref{table1},
supplemented by a (topological) theta term~\cite{Pruisken}.}

Let us now move on to $d=3$ dimensions, still remaining in symmetry class A.
Now the relevant homotopy group is (for sufficiently large values
of $n$ and $m$) 
$$
\Pi_{3} \left [
U(n+m)/[U(n)\times U(m)]
\right ]
= \{  {\bf 1} \},
$$
which is the trivial group of only one element,
as indicated.
This means that here band insulators
can only be in one phase. In $d=3$ spatial dimensions
there are hence no non-trivial topological
insulators in symmetry class A.

Are there then any topologically non-trivial band insulators in
$d=3$ dimensions at all? The 
answer%
\footnote{We have already mentioned in example (iii) of  the Introduction (Section A) 
that there exist in $d=3$ topological insulators in the presence of strong
spin-orbit interactions. In the language of Table~\ref{table1} and Table~\ref{table2}, these belong
to symmetry class AII, and will be discussed below.}
is ``yes''. We can see this for example from the observation
that the presence of SLS is a potential ``source'' of non-trivial topological behavior.
A look at Table~\ref{table1} reveals that there are five symmetry classes which
possess SLS, i.e., which have entries SLS$=1$. (This, as it turns out,
does not mean however, that there are non-trivial topological band 
insulators in all these five symmetry classes.) 
What is the technical benefit of SLS?
It arises from the observation\footnote{which is easy to check; see~\cite{SchnyderRyuFurusakiLudwig2008}.}
that  the presence of this symmetry implies
that the Hamiltonian ${\cal H}$ can
be brought into block off-diagonal form,
i.e., that
\begin{equation}
Q({\vec k})
\ \ 
=
\pmatrix{ 0 & q({\vec k}) \cr
 q^\dagger({\vec k}) & 0  \cr},
\quad 
{\rm where } \ \ q({\vec k})  \ \ {\rm is \ unitary}.
\label{QBlockOffDiagonal}
\end{equation}
Consider now the simplest symmetry class with SLS$=1$,
which possesses no symmetry
other than SLS. This is symmetry class AIII (4th row of Table \ref{table1}).
Due to the lack of any additional symmetry constraint,
$q({\vec k})$ is an arbitrary unitary matrix,
which fully characterizes a phase of the band insulator
in this symmetry class.
Thus, as in the case of class A considered before,
we now need to investigate the homotopy group of
maps from the BZ into the group $U(m)$ of unitary matrices.
This homotopy group is non-trivial in $d=3$ dimensions,
$$
\Pi_3 \left [ U(m) \right ]
=
{\bf Z}.
$$
This means that in symmetry class AIII there exists a distinct band insulator 
for every integer, and band insulators characterized
by different integers cannot be adiabatically deformed into each other
without crossing a quantum phase transition.
For completeness, let us also give the explicit form of the integer
$\nu\left (q({\vec k}) \right )$,
as a functional of 
$q({\vec k})$ which characterizes the Hamiltonian:

$$
\nu\left (q({\vec k}) \right )
=
$$
\begin{equation}
=
\int_{BZ}
{
d^3{\vec k}
\over 24 \pi^2}
\ \ 
\epsilon^{\mu\nu\rho}
\ \
tr
\left [
(q^{-1}\partial_\mu q)(q^{-1}\partial_\nu q)(q^{-1}\partial_\rho q)
\right ],
\label{ExpressionNuq}
\end{equation}
where $\epsilon^{\mu\nu\rho}$ is the usual totally antisymmetric tensor
($\epsilon^{1 2 3} =+1$).

Having presented the appearance of topological properties for band insulators
in the two symmetry classes A and AIII, we will now briefly comment on how to extend this
to the other classes, even though we will use a different approach
to arrive at the classification scheme, to be
discussed in the next section. For the (five) symmetry classes with SLS$=0 $,
the (simplified) Hamiltonian $Q ({\vec k})$ will satisfy additional conditions.
For example, in symmetry class AII [mentioned in example (iii)],
the TRS condition has to be imposed which reads
\begin{equation}
 \sigma^y \ Q^*({\vec k}) \ \sigma^y =  Q(-{\vec k}).
\label{ConstraintAtwo}
\end{equation}
Even though in $d=3$ dimensions there
existed only a single phase in class A (where $Q({\vec k})$
was subject to no constraints),
the set of all Hamiltonians
satisfying the additional constraint \eqref{ConstraintAtwo}
turns out to consist of {\it two} phases (or sectors) which
cannot be continuously deformed into each other.
Similarly, for all other symmetry classes with SLS$=1$, 
there will be certain constraints on the matrices $q({\vec k})$,
which appeared in \eqref{QBlockOffDiagonal}.
For example, in symmetry class CI one turns out to have
$q^t(-{\vec k}) = 
q({\vec k})$. A list of these constraints for all ten symmetry classes
is provided 
in Table III of~\cite{SchnyderRyuFurusakiLudwig2008}.

\section{D. Classification of $d=3$ topological insulators (superconductors)}

\begin{figure}
  \includegraphics[height=.25\textwidth]{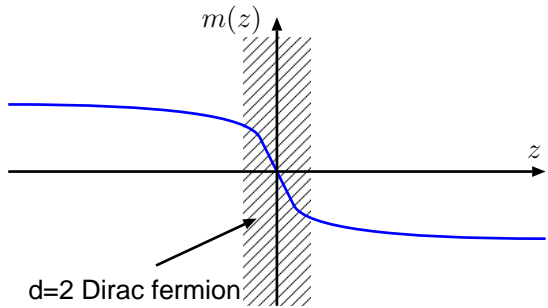}  
  \caption{
Domain wall arising form the change of sign of Dirac mass term.}
\label{FigureDomainWall}
\end{figure}

In this section we review the classification of 
$d=3$ topological insulators (superconductors).
This provides the main result of the work in 
\cite{SchnyderRyuFurusakiLudwig2008},
namely the last column of Table 2.
Our approach
is the one already mentioned in the Introduction (Section A).
We focus on the robustness of the gapless boundary (surface)
degrees of freedom:
for every  {\it topological} $d=3$ bulk  insulator (superconductor)
in one of the ten symmetry classes of Hamiltonians,  listed in Table 1,
there appear gapless degrees of freedom at its boundaries. These gapless
boundary degrees of freedom cannot be gapped or localized by any perturbations
or deformations of the Hamiltonian, whether these are (i) spatially
uniform or whether they (ii) break translational invariance
(i.e., are ``random''),
as long as these perturbations preserve
the symmetries of the given  symmetry class (Table 1). 
Our approach thus consists in going through the ten symmetry classes
of Hamiltonians in Table 1, {\it in $d=2$ dimensions} (describing
the boundary degrees of freedom),
and checking
whether localized or gapped boundary degrees of freedom are possible or not  in each class.
If localized or gapped boundary degrees of freedom are {\it not}
possible, then there exists a $d=3$ topological insulator (superconductor)
in this symmetry class (possessing these gapless boundary degrees of freedom).

In this analysis one needs to recognize the importance of one extra
ingredient: it is well known that $d=3$  massive Dirac Hamiltonians
possess topological properties; more specifically, when  changing the {\it sign}
of the Dirac mass term by letting that mass vary between positive and negative
values, say, in one 
direction
 (e.g., in the $z$ direction,
so that the mass is $m(z)$ as sketched in Fig.~\ref{FigureDomainWall}),
a gapless $d=2$ Dirac fermion degree of freedom will appear at the
``domain wall'' where the mass goes through 
zero~\cite{CallanHarvey,HaldanePRL1988,LudwigFisherShankarGrinstein1994}.
This shows that in general
one needs to allow for the Hamiltonian of the boundary degrees of freedom
to be of Dirac form (we will discuss this shortly in somewhat more detail below).
This is important because it was recently demonstrated by 
Bernard and LeClair~\cite{BernardLeClairJPHYSA2001}
that there are exactly 13 and not just  10 symmetry classes of
Dirac Hamiltonians in $d=2$ dimensions. 
This is due to the fact
that a  $d=2$ Dirac Hamiltonian has a special  $2 \times 2 $ block structure
(see \eqref{TwoDimensionalDiractHamiltonian} below),
allowing for a ``fine structure'' of the general classification of Table~\ref{table1}.
Bernard and LeClair's result thus means that some of the ten 
symmetry classes
from Table 1 subdivide into subclasses, and this is important for our
discussion.

In short, a $d=2$ Dirac Hamiltonian is of the form
\begin{equation}
{\cal H}
=
\pmatrix{
{\bf V}_+ + {\bf  V}_- & -i 
{\scriptstyle \partial\over \partial {\bar z}} 
{\bf 1}
 + {\bf A}_+ \cr
+i
 {\scriptstyle \partial\over \partial  z} 
{\bf 1}
 + {\bf A}_- & {\bf V}_+ - {\bf  V}_-  \cr} ,
\label{TwoDimensionalDiractHamiltonian}
\end{equation}
where 
$z=x+iy, {\bar z} = x-i y$ represent the $d=2$-dimensional spatial coordinates,
and ${\bf A_+}^\dagger={\bf A_-}$, and $ {\bf V_{\pm}}^\dagger= {\bf V_{\pm}}$
are $M\times M$ matrices, which are in 
general\footnote{for non-homogeneous (``random'') systems.}
functions of $(x,y)$ 
[here ${\bf 1}$ is the unit matrix].
We will refer to $M$ as the number of flavors of $d=2$ Dirac fermions.

Now, the findings of
Bernard and LeClair~\cite{BernardLeClairJPHYSA2001}
are easy to 
state:\footnote{For more details see the third column of 
Table III in Ref.~\cite{SchnyderRyuFurusakiLudwig2008}.}
in $d=2$ dimensions
there are 13  symmetry classes of Dirac Hamiltonians
\eqref{TwoDimensionalDiractHamiltonian}
because the three symmetry classes AIII, DIII and CI from Table~\ref{table1} subdivide
each into {\it two} subclasses.

Since it may be useful for some readers to
see the relevance of the Bernard-LeClair classification
in the case of the  ``quantum spin Hall'', or ``$Z_2$-topological insulator''
in $d=3$ dimensions [example (iii) from the Introduction, Section A],
 we will briefly review this connection in the following
subsection. In the subsequent subsection we will 
discuss all other symmetry classes.

\subsection{ D.1
$Z_2$-topological insulator in $d=3$ ($d=3$ version of the ``quantum
spin Hall'' state)
}

\begin{table}[th]
\label{table4}
\begin{tabular}{|c||c|c|c||c|c|}
\hline
Cartan nomenclature & TRS&PHS &SLS & Bernard-LeClair     & $M=$ \# of fermion species \\\hline\hline
AIII                & 0  &0   & 1  &$({\rm AIII})_{o}$   & $(2m-1)$ \\ 
                    &   &   &   &$({\rm AIII})_{e}$   & $2m$ \\  \hline \hline
DIII                & -1 &+1  & 1  &$({\rm DIII})_{o}$   & $(2m-1)$ \\ 
                    &    &  &   &$({\rm DIII})_{e}$   & $2m$ \\  \hline \hline
CI                  & +1 &-1  & 1  &$({\rm CI})_{o}$   & $(2m-1)\cdot 2$ \\ 
                    &    &  &   &$({\rm CI})_{e}$   & $ (2m)\cdot 2$ \\ \hline \hline
\end{tabular}
\caption{
Subdivision of Symmetry Classes AII, DIII and CI for $d=2$ Dirac Hamiltonians.
}
\end{table}

The work of Bernard and LeClair~\cite{BernardLeClairJPHYSA2001}
tells us\footnote{See, e.g., third column of 
Table III in Ref.~\cite{SchnyderRyuFurusakiLudwig2008}, or
Eq.\ (2.19) of \cite{BernardLeClairJPHYSA2001}.}
that there exists a $d=2$ Dirac Hamiltonian in symmetry class AII (Table 1)
with only a {\it single} flavor $M=1$
(in general, an odd number $M$) of 
Dirac fermions.
It is known that a {\it single} flavor cannot be realized in a $d=2$ lattice
model (due to the familiar ``fermion doubling'' phenomenon).
Therefore, this situation must correspond to the boundary of a $d=3$
topological insulator in symmetry class AII. Recall that this symmetry
class refers to the presence of a time-reversal symmetry
whose (anti-unitary) time-reversal operator squares to $-1$,
and is relevant for systems possessing spin-orbit coupling.
Indeed, a {\it single} flavor Dirac
fermion was constructed
explicitly by Fu, Kane, and Mele~\cite{REF3DZ2TopIns-Fu}
at the $d=2$ boundary of a three-dimensional (quantum spin Hall)
$Z_2$ topological insulator.
Bernard and LeClair show that the most general $d=2$ Dirac Hamiltonian
with a {\it single} flavor is a $2\times 2$
 matrix of the form
\begin{equation}
{\cal H}
= (-i) [
 \sigma_x \, \partial_x +
 \sigma_y \, \partial_y
]
+
V(x,y) {\bf 1} ,
\label{SingleFlavorDirac}
\end{equation}
where $V(x,y) $ is a ``scalar potential''.
It has long been 
known \cite{LudwigFisherShankarGrinstein1994,AndoEtAL1998}
that this Hamiltonian lies in symmetry class AII of Table~\ref{table1}.
Recent work established that this Hamiltonian cannot lead to localized
states~\cite{RyuEtAl2007,NomuraEtAl2007,OstrovskyEtAl2007,BardasonEtAl2007}:
indeed, in the presence of a random scalar potential
$V(x,y) $ the system behaves at large  length scales always
like a simple diffusive metal.

\subsection{ D.2 Topological insulators (superconductors) in $d=3$: all cases}

In the previous subsection we have seen that there exists
a topological insulator
in symmetry class AII in $d=3$ dimensions, because the boundary degrees of
freedom cannot be localized. In the presence of the only possible disorder
potential in this symmetry class, the system becomes the simplest
possible disordered metallic conductor.

Let us now turn our attention to the three symmetry classes AIII, DIII and CI
which, as already mentioned above,
subdivide into two subclasses each, when $d=2$  Hamiltonians
with a Dirac structure \eqref{TwoDimensionalDiractHamiltonian}
are considered.
As summarized in Table 3
these two subclasses simply correspond to whether
the number $M$ of flavors 
is even or odd.
(More precisely this is the case for  the two symmetry classes AIII and DIII; 
on the other hand, for the
time-reversal invariant Hamiltonians in class CI, the 
number of flavors $M$
is an even or an odd number of Kramers doublets, so that
$M=(2n-1)\cdot 2$ or $=2n\cdot 2$.)
As it turns out, the symmetry constraints in the special symmetry classes
$({\rm AIII})_{o}$, $({\rm CI})_{o}$, and $({\rm DIII})_{o}$ 
force~\cite{BernardLeClairJPHYSA2001} 
both potentials ${\bf V}_{\pm}$
in \eqref{TwoDimensionalDiractHamiltonian} to vanish identically.
Only the potentials ${\bf A}_{\pm}$ can be non-vanishing:
these, on the other hand, are nothing but non-Abelian gauge potentials
in the three classical groups (unitary, orthogonal, and symplectic,
for 
$({\rm AIII})_{o}$,$({\rm DIII})_{o}$, and $({\rm CI})_{o}$, respectively). 
Let us summarize this result:
\begin{equation}
\matrix{
U(2m-1)     &            & ({\rm AIII})_{o} \cr
SO(2m-1)    & {\rm for}  &  ({\rm DIII})_{o}\cr
Sp[2(2m-1)] &            & ({\rm CI})_{o}\cr
} \; \; .
\label{ClassicalGaugeGroups}
\end{equation}
Now, the important physical consequence of the fact that
the symmetries in classes
$({\rm AIII})_{o}$,$({\rm DIII})_{o}$, and $({\rm CI})_{o}$, 
allow only for the presence of gauge potentials 
is that such potentials, whether homogeneous or random,
cannot (see, e.g., 
~\cite{LudwigFisherShankarGrinstein1994,Tsvelik1995,Mudry1996,Bhasen2001,Ludwig2000}) 
localize or gap out
the Dirac fermions (which are
certainly gapless in the absence of any potentials).
The behavior of $d=2$ Dirac fermions in the presence
of these random potentials is a 
well-studied problem
(see, e.g., 
~\cite{LudwigFisherShankarGrinstein1994,Tsvelik1995,Mudry1996,Bhasen2001,Ludwig2000}):
even though disorder may lead to highly non-trivial and interesting
behavior\footnote{see, e.g.,~\cite{SchnyderRyuLudwig2009arXiv} 
for the example
of the topological superconductor in symmetry
class CI; Ref.~\cite{SchnyderRyuLudwig2009arXiv} 
includes also
a corresponding  brief discussion of the cases AIII and DIII.
}, 
the value of the  longitudinal surface
conductivity\footnote{thermal ($\kappa_{xx}/T$) or, if SU(2) spin
rotation symmetry is preserved by the Hamiltonian, 
spin conductivity $\sigma^{spin}_{xx}$ for
superconductors.}
is {\it unchanged} by this type of disorder.
This means that irrespective of
the presence of disorder,
the
longitudinal surface conductivity
$\sigma_{xx}$
is ${1\over \pi} (e^2/h)$ times the number of Dirac fermion flavors $M$.
(In the case 
of spin- or thermal conductivity,
 $\sigma^{spin}_{xx}$ or $\kappa_{xx}/T$,
the conductance unit $(e^2/h)$ has of course to be replaced
by  the corresponding unit~\cite{SchnyderRyuFurusakiLudwig2008}.)
The number $M$ of 
Dirac fermion flavors is directly related to the integer-valued
topological ``winding number'' $\nu$, discussed in
\eqref{ExpressionNuq}
in the spatially
homogeneous case. Thus, the longitudinal surface conductance
is a direct measure of the topological index of the bulk
of the topological insulator (superconductor) in these
symmetry classes.

It remains to discuss symmetry class CII. One can show~\cite{SchnyderRyuFurusakiLudwig2008}
that in this class of Dirac Hamiltonians neither spatially homogeneous nor
inhomogeneous (random) potentials can 
gap out or localize the
$d=2$ surface degrees of
freedom {\it if} the number of Dirac fermion species 
is an odd multiple of two, corresponding to an odd number of Kramers
doublets (this class possesses time reversal symmetry).
One can also show that this corresponds only to a $Z_2$ classification,
corresponding to $M=0$
(topologically trivial) and $M=2$
(topologically non-trivial); changing
$M$ by four (i.e., by two Kramers doublets)
does not lead to a topologically different state.

Finally, it is very easy to see~\cite{SchnyderRyuFurusakiLudwig2008}
from the Bernard-LeClair classification that
in all the remaining symmetry classes of the 10 classes
(i.e., in classes A, AI, BDI, D, C)
the $d=2$ Dirac Hamiltonian
can be made fully gapped while keeping all defining
symmetries of the  class intact.
By performing these steps,
we arrive at the results listed in the  last column (entitled ``$d=3$'')
of Table \ref{table2}.
This is our main result,
obtained in \cite{SchnyderRyuFurusakiLudwig2008}.

\begin{table}[th]
\label{table3}
\begin{tabular}{|c||c|c|c||c|c|c|c|}
\hline
   Cartan nomenclature & TRS & PHS & SLS & Hamiltonian &  $d=1$     & $d=2$       & $d=3$           \\\hline\hline
     AIII (chiral unit.)   &$0$  &$0$  &$1$ &${\scriptstyle U(N+M)/U(N)\times U(M)}$  &${\bf Z}$  &  -          & ${\bf Z}$      \\ \hline
    A (unitary)         & $0$ &$0$  & $0$ & $\scriptstyle U(N)$&  -             & ${\bf Z}$   &  -            \\ \hline\hline
BDI (chiral orthog.)  &$+1$ &$+1$ &$1$ & $\scriptstyle SO(N+M)/SO(N)\times SO(M)$  &${\bf Z}$  & -           &        - \\ \hline
D                      &$0$  &$+1$ &$0$ & $\scriptstyle SO(2N)$                    &${\bf Z}_2$& ${\bf Z}$   & - \\ \hline
DIII                   &$-1$ &$+1$ &$1$ & $\scriptstyle SO(2N)/U(N)$          &${\bf Z}_2$&${\bf Z}_2$  & ${\bf Z}$ \\ \hline
AII (symplectic)      &$-1$ &$0$  &$0$ & $\scriptstyle U(2N)/Sp(2N)$  &  -        &  ${\bf Z}_2$& ${\bf Z}_2$    \\\hline
CII (chiral sympl.)   &$-1$ &$-1$ &$1$ & $\scriptstyle Sp(2N+2M)/Sp(2N)\times Sp(2M)$  &${\bf Z}$  & -           & ${\bf Z}_2$   \\\hline 
C                      &$0$  &$-1$ &$0$ & $\scriptstyle Sp(2N)$    & -         &  ${\bf Z}$  & -        \\ \hline
CI                     &$+1$ &$-1$ &$1$ & $\scriptstyle Sp(2N)/U(N)$  & -   & -    & ${\bf Z}$ \\ \hline
AI (orthogonal)       &$+1$ &$0$  &$0$ & $\scriptstyle U(N)/O(N)$  &  -        & -           &  -               \\ \hline
 \end{tabular}
\caption{
Reorganizing Table~\ref{table2} by
reordering the symmetry classes and grouping them into two separate 
lists reveals a regular pattern, which was recently pointed out by 
A.\ Kitaev Ref.~\cite{KitaevLandau100Proceedings}.
} 
\end{table}

\section{E. Classification of $d=2$ topological insulators (superconductors)}

We briefly summarize from \cite{SchnyderRyuFurusakiLudwig2008} the classification
of $d=2$ topological insulators (superconductors). It may
be useful for the reader to follow the discussion by keeping
an eye on Table~\ref{table2}. There are three well known symmetry classes
which support topological insulators (superconductors)
in $d=2$: these are symmetry classes A, D, and C,
all of which break time-reversal symmetry, and are all known~\cite{ReadGreen,Prange1990,RefsTQHE,RefsSQHE,Senthil1998}
to possess a quantum Hall insulating state. The latter manifests
itself by the appearance of chiral edge states.
Classes A and D were discussed in examples (i) and (ii) in the Introduction (Section A),
and class C is known as the so-called spin quantum Hall effect~\cite{RefsSQHE,Senthil1998} 
(not to be confused with the ``quantum spin Hall state'' discussed in example (iii)
of the Introduction). Since these states may possess any number of chiral edge
states, the different topological sectors
of the $d=2$ insulators (superconductors)
are characterized by integers. This is the origin of the entries ${\bf Z}$
in the penultimate column of Table~\ref{table2}. This same column contains
in addition 
an entry $Z_2$ in the  row labeled AII: this is
the $d=2$ ``quantum spin Hall''
insulator discussed in example (iii) of the Introduction.
It remains to discuss the row labeled DIII.
This case was treated in \cite{GruzbergReadVishveshwara},
where the authors studied the localization physics of (quasi-) one-dimensional
systems: the authors found that a  
\hbox{(quasi-)} one-dimensional Hamiltonian in symmetry class DIII
cannot be localized or gapped
if there is an odd number of 
one-dimensional
modes. This situation
can be realized~\cite{SchnyderRyuFurusakiLudwig2008,Roy08,QiZhang08} 
in chiral p-wave superconductors
with opposite chiralities
[$(p_x+ip_y)$ and $(p_x-ip_y)$ pairing symmetries].
Moreover, 
in the remaining five symmetry classes of Table~2,
(quasi-) one-dimensional Hamiltonians will
always generically be  localized or gapped.
This is related, in great generality, to the
well-known ``12-fold way''
classification scheme\footnote{Two of the ten symmetry classes
undergo a subdivision; these are precisely
the symmetry classes AII and DIII where $Z_2$ topological
insulators exists in $d=2$. The existence of  
these topological states is precisely related to this splitting.
This is  similar to what happened in the Bernard LeClair
classification scheme~\cite{BernardLeClairJPHYSA2001}
for $d=2$ Dirac Hamiltonians, discussed in Section D.}
of random transfer matrices,
summarized in Table IV of \cite{SchnyderRyuFurusakiLudwig2008}.

\section{F. Classification of $d=1$ topological insulators (superconductors)}

Again, we proceed~\cite{SchnyderRyuFurusakiLudwig2008}
as for dimensions
$d=2$ and $d=3$, reviewed above:
the diagnostic of a $d=1$ topological insulator (superconductor)
is the appearance of
gapless degrees of freedom at the boundaries. In $d=1$ the boundaries
are points.
Thus, we need to check in which of the 10 symmetry classes
of Table~\ref{table1} gapless degrees of freedom (``zero modes'')
appear at a point.
The answer to this question is known from random matrix 
theory\footnote{
describing Hamiltonians in spatial dimension $d=0$ 
},
and was found for all 10 symmetry classes in 2001 by D. Ivanov~\cite{Ivanov}.
A summary of these results is displayed in Table V of
\cite{SchnyderRyuFurusakiLudwig2008}. Using this information,
one arrives at the column entitled ``$d=1$'' of Table~\ref{table2}.

\section{G. Discussion}

Table~\ref{table2} summarizes the main result of this work, the classification of
topological insulators (superconductors) in spatial dimension $d=1,2,$ and 3.%
\footnote{
One can understand the presence of the
key signatures of  topological insulators, 
namely, the stability of their gapless nature and
the complete absence of Anderson localization for boundary degrees of freedom,
using a variety of techniques and from different points of views.
The following presents yet another slightly different way of
thinking about this.
The Anderson localization problem at the boundary of topological insulators
can also be discussed in terms of the 
NLSM formalism in a rather unified fashion. (We may choose here the
so-called ``fermionic replica'' formulation
(see the last column in Table~\ref{table2}),
but  we may, equivalently, choose the formulation
using supersymmetry~\cite{EfetovBook}.)
When the NLSM formalism is applied to describe effects of disorder
on the gapless boundary degrees of freedom,
the fact that
these boundary degrees of freedom
completely evade Anderson localization
is signaled by an
additional term which can be added to the NLSM action.
Depending on the symmetry class and spatial dimensions,
it takes on the form of either a topological or
a Wess-Zumino-Witten (WZW) term.
In turn, the presence or absence of 
a topological or WZW term
for a given symmetry class in $d$ dimensions
can be read off 
from Bott periodicity -- see below. (Compare also footnote 9.)
}
The symmetry classes in Table~\ref{table2} are organized
according to the physical systems these symmetry classes represent (three
Wigner-Dyson classes of standard electronic systems; three Wigner-Dyson
classes with extra (``sublattice'' or ``chiral'' symmetry: SLS$=1$);
four classes of BdG Hamiltonian in superconductors).
While such an ordering is natural from the physics point of view,
it hides an underlying mathematical structure, namely
a periodicity in spatial dimension $d$, which was
recently pointed out by Kitaev~\cite{KitaevLandau100Proceedings}.
To uncover this periodicity we have reorganized Table~\ref{table2} by
reordering the symmetry classes and grouping them into two separate 
lists (see Table~\ref{table3}).
The upper list in Table~\ref{table3} contains
only the two
unitary classes A and AIII, both of which 
have neither TRS nor PHS (discussed in Sec.\ B).
These two classes are related
to the two types 
of classifying spaces
appearing in {\it complex} K-Theory
of Ref.~\cite{KitaevLandau100Proceedings}.
We see from the reordered Table~\ref{table3}
that an
alternating pattern (period 2) in the spatial dimension $d$ becomes apparent:
i.e., the class AIII topological insulator can only exist
in odd spatial dimensions,
while the class A topological insulator
(i.e., the integer quantum Hall insulator)
occurs only in even spatial dimensions.
The lower list in Table~\ref{table3} (classes BDI, D, $\cdots$, AI) contains
all the remaining classes; those are the ones
that have at least either TRS
or PHS. These eight classes are related to the eight types
of classifying spaces
appearing in {\it real} K-Theory,
discussed in Ref.~\cite{KitaevLandau100Proceedings}.
An obvious regular pattern emerges when looking at the reordered
Table~\ref{table3}:
as the spatial dimension $d$
is increased by one,  the topological
insulators (superconductors) move down by one column.
It was shown by Kitaev  that this regular pattern
is due to an $8$-fold 
periodicity in $d$, the 
Bott periodicity of real K-theory.  
Taking this result from K-theory, we can extend our result to
dimensions $d>3$.
For example, Table~\ref{table3} suggests that in $d=4$ there is a 
topological insulator
whose topologically distinct sectors are classified by integers ${\bf Z}$,
which belongs
to symmetry class AII. 
Indeed, Qi \textit{et al.} \cite{ShouChengEtAl} have recently 
shown that the $Z_2$ topological insulators of the class AII in $d=2$ and 3
can be obtained as descendants from this four-dimensional $Z$ topological
insulator using dimensional reduction.


\begin{theacknowledgments}

We thank Alexei Kitaev for stimulating discussions. 
This work was supported in part by the National Science Foundation (NSF)
under Grant Numbers PHY05-51164 (S.R., A.S., A.F.)
and DMR-0706140 (A.W.W.L.). A.W.W.L. thanks the organizers of the 
{\it Landau Memorial Conference ``Advances in Physics''} 
for the opportunity to present this work.

\end{theacknowledgments}


\end{document}